\documentclass[pra,reprint,superscriptaddress]{revtex4-1}
\pdfoutput=1
\usepackage{amsmath,amssymb, graphicx, hyperref, natbib, color, lipsum}

\newcommand{\+}{^\dagger}
\newcommand{\nodag}{^{\phantom{\dagger}}}
\newcommand{\ha}{\hat{a}}

\newcommand{\hH}{\hat{H}}

\newcommand{\expect}[1]{\left\langle #1 \right\rangle}
\newcommand{\hsigma}{\hat{\sigma}}
\newcommand{\determinant}[1]{\left|\begin{array}{cccc}#1 \end{array}\right|}

\begin{document}
\author{Bhuvanesh Sundar}
\email{bs55@rice.edu}
\affiliation{Department of Physics and Astronomy, Rice University, Houston, TX 77005}
\affiliation{Rice Center for Quantum Materials, Rice University, Houston, Texas 77005, USA}

\author{Marc Andrew Valdez}
\email{mvaldez@mymail.mines.edu}
\affiliation{Department of Physics, Colorado School of Mines, Golden, CO 80401}

\author{Lincoln D. Carr}
\email{lcarr@mines.edu}
\affiliation{Department of Physics, Colorado School of Mines, Golden, CO 80401}

\author{Kaden R. A. Hazzard}
\email{kaden@rice.edu}
\affiliation{Department of Physics and Astronomy, Rice University, Houston, TX 77005}
\affiliation{Rice Center for Quantum Materials, Rice University, Houston, Texas 77005, USA}

\title{A complex network description of thermal quantum states in the Ising spin chain}
\date{\today}

\begin{abstract}
We use network analysis to describe and characterize an archetypal quantum system -- an Ising spin chain in a transverse magnetic field. We analyze weighted networks for this quantum system, with link weights given by various measures of spin-spin correlations such as the von Neumann and R\'enyi mutual information, concurrence, and negativity. We analytically calculate the spin-spin correlations in the system at an arbitrary temperature by mapping the Ising spin chain to fermions, as well as numerically calculate the correlations in the ground state using matrix product state methods, and then analyze the resulting networks using a variety of network measures. We demonstrate that the network measures show some traits of complex networks already in this spin chain, arguably the simplest quantum many-body system. The network measures give insight into the phase diagram not easily captured by more typical quantities, such as the order parameter or correlation length. For example, the network structure varies with transverse field and temperature, and the structure in the quantum critical fan is different from the ordered and disordered phases.
\end{abstract}
\maketitle

\section{Introduction}
Network analysis is a powerful technique to characterize the structure of connections between agents in a network~\cite{newman2003structure, bornholdt2006handbook}. Studies have shown that classical systems as diverse as the brain and the Internet have a complex network structure~\cite{papo2014complex, bullmore2009complex, rubinov2010complex, telesford2011brain, khan2017emerging, carmi2007model}. Quantum systems also show a wide variety of complexity emerging due to inter-particle interactions. Like classical systems, quantum systems have an interconnected web of correlations, and network analysis provides a powerful set of tools to study them. However, while complex networks are ubiquitous in classical systems with a sufficiently rich set of interacting components, it is an open question what the minimal interacting quantum many-body system is in which complex network structures can appear.

In this paper, we address this question by studying the network of correlations that arises in the simplest of interacting quantum models, the one-dimensional transverse field Ising model (TIM). We introduce and calculate networks whose links are weighted by various measures of correlations and entanglement, and quantify their complexity. The emergence of network complexity illuminates the richness of the quantum system.

Earlier works have studied complex networks in the context of quantum systems, but by enforcing complex network structure in the Hamiltonian, e.g, in interactions~\cite{Bianconi2001, Bianconi2012a, Bianconi2012b, Kimble2008, Perseguers2010, Cuquet2009}. However, there is no need for this explicit enforcement, as one finds network structure already in quantum states even for simple models such as the nearest-neighbor TIM. The network naturally arises in their correlations, just as it does in classical systems. We quantify the network's complexity via network measures. Quantifying this complexity at zero temperature has found applications such as identifying phase transitions~\cite{valdez2017quantifying}.

We calculate networks for the spin-spin correlations, the mutual information, concurrence, and negativity of spins, using analytical solutions for the two-spin reduced density matrix at an arbitrary temperature and magnetic field in the thermodynamic limit. We also numerically calculate these networks for the ground state of a finite system, using matrix product state (MPS) methods~\cite{schollwock2005density, verstraete2008matrix, schollwock2011density} implemented in the openMPS code of Ref.~\cite{jaschke2017open}. All these networks are calculable from measurements in a variety of experiments on cold atoms and trapped ions~\cite{roos2004bell, haffner2005scalable, cramer2010efficient, lanyon2017efficient, parsons2016site, cheuk2015quantum, edge2015imaging, omran2015microscopic, haller2015single, yamamoto2016ytterbium, richerme2014non, zhang2017observation, zeiher2016many, greif2015formation, bernien2017probing}. We analyze the structure of these networks, specifically their density, disparity, betweenness-centrality, clustering coefficient, average geodesic distance, and diameter. [We define these network measures in Sec.~\ref{subsec: network measures}.]

This article is organized as follows. In Sec.~\ref{sec: networks}, we describe the networks and network properties we use to characterize quantum systems. In Sec.~\ref{sec: TIM}, we describe the TIM, and its analytical and numerical solutions. In Sec.~\ref{sec: network analysis for TIM}, we calculate network measures for the networks described in Sec.~\ref{sec: networks}. We conclude in Sec.~\ref{sec: conclusions}.

\section{Complex networks}\label{sec: networks}
\begin{figure}[t]
\centering
\includegraphics[width=0.7\columnwidth]{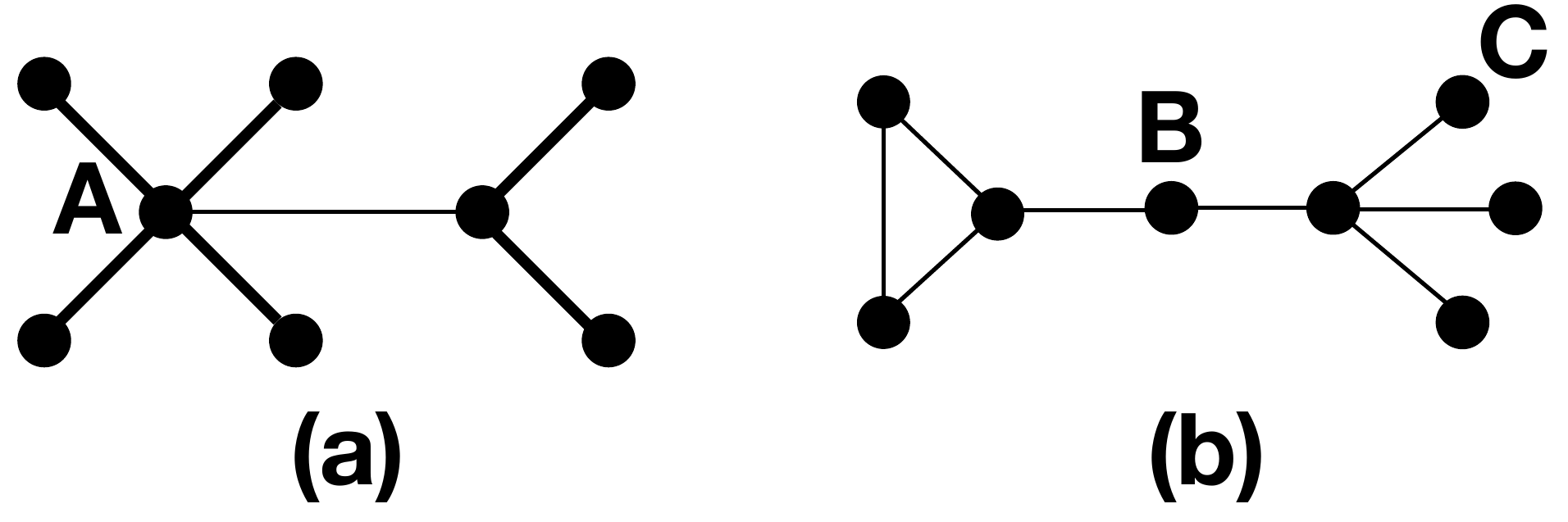}
\caption{Examples of weighted networks illustrating the meanings of the different network measures we use, namely the \textit{local measures} density, disparity, and betweenness-centrality for a node, and the \textit{global measures} clustering coefficient, average geodesic distance, and diameter for a network. In the examples above, the thickness of a link is proportional to its weight. Node $A$ has a large density and betweenness-centrality, and a small disparity. Node $B$ has a small density and disparity, and a large betweenness-centrality. Node $C$ has a small density and betweenness-centrality, and a large disparity. The network in (a) has a smaller diameter, average geodesic distance, and clustering coefficient than the network in (b).}
\label{fig: network examples}
\end{figure}

Complex networks are networks with non-trivial features in their topology and connectivity that are not usually found in other networks such as lattices or random graphs~\cite{newman2003structure}. Natural and social communities furnish abundant examples of complex networks, e.g. the Internet, social media, citation networks, neural networks in the brain, food webs, and so on (see \cite{strogatz2001exploring} and references therein). These networks come in many kinds: they can be dense, disparate, uniform, or clustered. For example, food webs are dense, metabolic networks are disparate, and social networks are highly clustered \cite{williams2000simple, jeong2000large, girvan2002community}.

For interacting quantum spins, we define an undirected weighted network to represent the system, where node $i$ represents the $i^{\rm th}$ spin, and link weight $e_{ij}$ is some measure of correlation between spins $i$ and $j$. The network is defined to have no self-connections, $e_{ii}=0\ \forall i$. Even short-range interactions between spins lead to complex long-range correlations in the system, leading to a rich network structure that varies across the phase diagram.

Our main theme is that the networks that we define effectively represent information about the quantum system. We demonstrate this by using network analysis techniques. We show that the variation of the network measures with system parameters -- magnetic field and temperature -- mirrors the underlying phase diagram.

\subsection{Network measures}\label{subsec: network measures}
Network measures quantify the distinguishing features of networks. These measures can be defined locally on a node, or globally for a network. Commonly used local measures for nodes are the  density $d_i$, disparity $Y_i$, and various centrality measures such as the betweenness-centrality $B_i$, the eigenvector centrality $E_i$, and the Katz centrality $K_i$. Some global network measures are the clustering coefficient \text{C}, average geodesic distance $\overline{D}$, and the diameter $D_{\rm max}$. We define these network measures as applicable to weighted networks. For a network with $N$ nodes and link weights $e_{ij}$ between nodes $i$ and $j$, the network measures we use are defined as
\begin{equation}\label{eqns: network measures}
\begin{array}{lllr}
\begin{array}{l}\text{Density:}\end{array} & d_i &= &\frac{\sum_j e_{ij}}{N-1},\\
&&&\\
\begin{array}{l}\text{Disparity:}\end{array} & Y_i &= &\frac{\sum_j e_{ij}^2}{\left(\sum_j e_{ij}\right)^2},\\
&&&\\
\begin{array}{l}\text{Betweenness-}\\ \text{centrality:}\end{array} & B_i &= &\sum_{j,k\neq i} \frac{N_{jik}}{N_{jk}},\\
&&&\\
\begin{array}{l}\text{Clustering}\\ \text{coefficient:}\end{array}& \text{C} &= &\frac{\sum_{i\neq j\neq k} e_{ij}e_{jk}e_{ki}}{\sum_k\sum_{i\neq j\neq k} e_{ik}e_{jk}},\\
&&&\\
\begin{array}{l}\text{Average\ geodesic-}\\ \text{distance:}\end{array} &\overline{D} &= &\frac{\sum_{ij} D_{ij}}{N(N-1)/2},\\
&&&\\
\begin{array}{l}\text{Diameter:}\end{array} & D_{\rm max} &= &\underset{i,j}{\rm max}\ D_{ij},
\end{array}\end{equation}
where $D_{ij}$ is the geodesic distance (i.e, length of the shortest path) between nodes $i$ and $j$, with the distance of a direct path from $i$ to $j$ defined as $1/e_{ij}$. $N_{jk}$ is the number of geodesic paths from $j$ to $k$, and $N_{jik}$ is the number of geodesic paths from $j$ to $k$ via $i$. We describe these measures by calculating them for the network examples depicted in Fig.~\ref{fig: network examples}. The density $d_i$ quantifies the importance of node $i$ as the sum of the link weights connected to it; for example, in Fig.~\ref{fig: network examples}, $d_A>d_B>d_C$. The disparity $Y_i$ quantifies how dissimilar a node's connections are; in Fig.~\ref{fig: network examples}, $Y_A<Y_B<Y_C$. The betweenness-centrality $B_i$ measures the importance of node $i$ to the connectivity of other parts of the network to each other. A node has a high betweenness-centrality if removing it distances many other parts of the graph from each other. In Fig.~\ref{fig: network examples}, $B_A=B_B>B_C$. The clustering coefficient of a network measures cohesiveness of the network. For an unweighted network, the clustering coefficient is thrice the ratio of the number of triangles (three mutually connected vertices) in the network to the number of triplets (three connected vertices). For example, the network in Fig.~\ref{fig: network examples}(a) has no triangles, and 12 triplets. The network in Fig.~\ref{fig: network examples}(b) has one triangle formed by the three leftmost nodes, and 12 triplets. We have generalized the definition of the clustering coefficient to apply to a weighted network in Eq.~\eqref{eqns: network measures}. The diameter $D_{\rm max}$ is the geodesic distance between the most distant pair of nodes, and the $\overline{D}$ is the average geodesic distance between all pairs of nodes. The network in Fig.~\ref{fig: network examples}(a) has a smaller diameter, average geodesic distance, and clustering coefficient, than the network in Fig.~\ref{fig: network examples}(b).

The networks that we consider in the rest of this paper differ from the examples in Fig.~\ref{fig: network examples} in at least two important respects: we mostly consider the thermodynamic limit, and all nodes are identical due to translational symmetry. Although correlations are calculated in the thermodynamic limit, in practice we truncate the size of the graph to $N\sim\mathcal{O}\left(100\right)$. We systematically analyze the convergence of our network measures as $N$ increases (see Figs.~\ref{fig: TIM_MI_openMPS}, ~\ref{fig: finite size scaling}, ~\ref{fig: TIM_MIRenyi_openMPS}, ~\ref{fig: TIM_con}c, and ~\ref{fig: TIM_con}d). The translational symmetry has a few consequences. First, it is meaningful to define the density $d$, disparity $Y$, and the betweenness-centrality $B$ for the network as the density, disparity, and betweenness-centrality of an arbitrarily chosen node. Second, some network measures are simply related to others. For example, the eigenvector centrality $E_i$ and Katz centrality $K_i$, which are defined as the solutions to:
\begin{align}
\sum_j e_{ij}E_j &= \lambda E_i,\nonumber\\
\alpha\sum_j e_{ij}(K_j+1) &= K_i,
\end{align}
with $\lambda$ the largest eigenvalue of the adjacency matrix with matrix elements $e_{ij}$, and $\alpha$ an arbitrary real number between $0$ and $1$, are both trivially related to the network measures already defined in Eq.~\eqref{eqns: network measures}. Translational symmetry, and the fact that all link weights are positive, imply that the eigenvector centrality of all nodes is $E_i=1$, and $\lambda = Nd_i$. The quantity $\lambda$ is called the strength of the network. The Katz centrality immediately follows as
\begin{equation}
K_i = \frac{\alpha\lambda}{1-\alpha\lambda}.
\end{equation}
We do not explicitly present these measures since they follow immediately from the density $d_i$ for our networks due to translational invariance.

\subsection{Correlation and entanglement networks for the spin chain}\label{subsec: building the network}
We denote the magnetization $m_j^\mu$ of the $j^{\rm th}$ spin along the $\mu$ direction, and the correlations between spins $i$ and $j$, as
\begin{subequations}\begin{align}
m_j^\mu &= \expect{\hsigma_j^\mu},\\
c_{ij}^{\mu\nu} &= \expect{\hsigma_i^\mu\hsigma_j^\nu}.
\end{align}\end{subequations}
The reduced density matrices for one and two spins are
\begin{subequations}\begin{align}
\rho_j^{(1)} &= \frac{1}{2}\sum_{\mu=0}^3 m_j^\mu\sigma^\mu,\label{eqn: rho1}\\
\rho_{ij}^{(2)} &= \frac{1}{4}\sum_{\mu,\nu=0}^3c^{\mu\nu}_{ij}\sigma^\mu\otimes\sigma^\nu \label{eqn: rho2}
\end{align}\end{subequations}
respectively, where we denote $\left(\sigma^0,\sigma^1,\sigma^2,\sigma^3\right) = \left(\mathbf{1},\sigma^x,\sigma^y,\sigma^z\right)$. From these, we calculate several measures that describe correlations and identify entanglement between sites $i$ and $j$: the von Neumann mutual information $\mathcal{I}_{ij}$, the R\'enyi mutual information $\mathcal{I}^q_{ij}$, the concurrence $\mathcal{C}_{ij}$, and the negativity $\mathcal{N}_{ij}$. These correlation measures are defined as
\begin{subequations}\begin{align}
&\mathcal{I}_{ij} = \frac{1}{2}\operatorname{Tr}\left(\rho_{ij}^{(2)}\log_2\rho_{ij}^{(2)} - \rho_i^{(1)}\log_2\rho_i^{(1)} - \rho_j^{(1)}\log_2\rho_j^{(1)}\right),\label{eqn: MI}\\
&\mathcal{I}^q_{ij} = \frac{1}{2(1-q)}\log_2\left( \operatorname{Tr}\left(\rho_i^{(1)}\otimes \rho_j^{(1)}\right)^q/ \operatorname{Tr}\left(\rho_{ij}^{(2)}\right)^q \right),\label{eqn: Renyi}\\
&\mathcal{C}_{ij} = {\rm max}\left(0,\lambda_1^{ij}-\lambda_2^{ij}-\lambda_3^{ij}-\lambda_4^{ij}\right),\label{eqn: con}\\
&\mathcal{N}_{ij} = \frac{\operatorname{Tr}\left|(\mathbf{1}\otimes {\rm T})\rho_{ij}^{(2)}\right|-1}{2}, \label{eqn: neg}
\end{align}\end{subequations}
where $\lambda_1^{ij}, \lambda_2^{ij}, \lambda_3^{ij}, \lambda_4^{ij}$ are the eigenvalues of $\sqrt{\sqrt{\rho_{ij}^{(2)}} \tilde{\rho}_{ij}^{(2)} \sqrt{\rho_{ij}^{(2)}}}$ in decreasing order, $\tilde{\rho}_{ij}^{(2)}= \hsigma_i^y\hsigma_j^y (\rho_{ij}^{(2)})^* \hsigma_i^y\hsigma_j^y $ is the spin-flipped reduced density matrix, and ${\rm T}$ is the transpose operator.

For each type of correlation, we define a network whose links are weighted by that correlation. (For the negativity network, we define the link weights as $e_{ij} = -\mathcal{N}_{ij}$, to keep them positive). We analyze these networks using the measures defined in Eq.~\eqref{eqns: network measures}. These network properties provide a wealth of information about the underlying system. All six network measures -- density, disparity, betweenness-centrality, clustering coefficient, average geodesic distance, and the diameter -- or their gradients are observed to have extrema at quantum phase transitions. This appears to be true regardless of the correlation measure  -- von Neumann or R\'enyi mutual information, spin-spin correlation, concurrence, or negativity -- used to build the graph. Ref.~\cite{valdez2017quantifying} has also shown that network measures undergo sharp changes for a wide variety of zero-temperature phase transitions -- mean field, $Z_2$, and Berezinskii-Kosterlitz-Thouless.

\section{Transverse field Ising model and solutions}\label{sec: TIM}
The TIM for a one-dimensional chain of $L$ spins is
\begin{equation}
\hH_{\rm TIM} = \sum_j\left( -J\hsigma^z_j \hsigma^z_{j+1} + h \hsigma^x_j\right) \label{eqn: HTIM}
\end{equation}
where the sum runs over all spins. In Sec.~\ref{subsec: analytical}, we present analytical solutions for spin-spin correlations at an arbitrary temperature in the thermodynamic limit $L=\infty$. (We still truncate our correlation networks at $N\sim\mathcal{O}(100)$). In Sec.~\ref{subsec: numerical}, we complement these solutions with a zero-temperature numerical calculation of spin-spin correlations in finite $L$ systems using MPS methods.

\subsection{Analytical solution at nonzero temperature}\label{subsec: analytical}
In the thermodynamic limit, $\hH_{\rm TIM}$ can be diagonalized by mapping spin operators to fermionic annihilation and creation operators $\ha_j\nodag$ and $\ha_j\+$ via a Jordan-Wigner transformation~\cite{jordan1993paulische, lieb1961two}: 
\begin{align}
\hsigma^x_j &= 2\ha_j\+\ha_j\nodag-1\nonumber\\
\hsigma^y_j &= i(-1)^{\sum_{k<j}\ha_k\+\ha_k\nodag} \left(\ha_j\+ - \ha_j\nodag\right)\nonumber\\
\hsigma^z_j &= (-1)^{\sum_{k<j}\ha_k\+\ha_k\nodag} \left(\ha_j\+ + \ha_j\nodag\right).
\label{eqn: JordanWigner}
\end{align}
Under this transformation, $\hH_{\rm TIM}$ gets mapped to a Hamiltonian that describes the Kitaev chain~\cite{kitaev2001unpaired}:
\begin{equation}
\hH_{\rm JW} = \sum_j J\left(\ha_{j+1}\+ + \ha_{j+1}\nodag\right)\left(\ha_j\+ - \ha_j\nodag\right) + h \left(2\ha_j\+\ha_j\nodag -1\right). \label{eqn: HKitaev}
\end{equation}
We diagonalize $\hH_{\rm JW}$ by rotating it into the basis of non-interacting Bogoliubov quasiparticles~\cite{bogoljubov1958new, valatin1958comments}. $\hH_{\rm JW}$ is known to have a topologically non-trivial superconducting ground state at $T=0$ for $|h|<J$, and a trivial superconducting ground state at $T=0$ for $|h|>J$, corresponding to the ferromagnetic and paramagnetic phase in the TIM, respectively. When $T>0$, thermal fluctuations break long-range order, and the system is always in the paramagnetic phase.

Spin observables for the Ising system at temperature $T$ can be calculated from the thermal equilibrium state of $H_{\rm JW}$, together with the Jordan-Wigner transformation [Eq.~\eqref{eqn: JordanWigner}]. It is useful here to define a quantity
\begin{equation}
Q_n = \expect{(\ha_n\+ + \ha_n\nodag)(\ha_0\nodag - \ha_0\+)}.
\end{equation}
In the thermodynamic limit,
\begin{equation}
Q_n = -\int_{-\pi}^\pi \frac{dk}{2\pi}\ e^{ikn} \frac{2\left(h-Je^{-ik}\right)}{\omega_k} \tanh\frac{\beta\omega_k}{2},
\end{equation}
where $\beta=1/k_BT$, and $\omega_k = 2\left|h-Je^{-ik}\right|$. It is then straightforward to show that
\begin{equation}
\expect{\sigma^x_i} = Q_0.
\end{equation}
We consider a system that is in a superposition which does not break $Z_2$ spin symmetry. Therefore, $\expect{\sigma^y_i} = \expect{\sigma^z_i} = 0$. [This corresponds to choosing the fermionic system to be in a state with a fixed number parity.]

The product $\sigma^\mu_{i}\sigma^\nu_{i+n}$ can be written as a string of fermionic operators. Correlations between spins can then be decomposed and simplified using Wick's theorem. The result is:
\begin{align}
c^{xx}_{i,i+n} &= Q_0^2-Q_nQ_{-n},\nonumber\\ \nonumber \\
c^{yy}_{i,i+n} &= \determinant{Q_{-1} & Q_0 & \cdots & Q_{n-2}\\ Q_{-2} & Q_{-1} & \cdots & Q_{n-3} \\ \vdots & \vdots & \ddots & \vdots\\Q_{-n} & Q_{-n+1} & \ldots & Q_{-1}}, \nonumber\\ \nonumber \\
c^{zz}_{i,i+n} &= \determinant{Q_1 & Q_2 & \cdots & Q_n\\ Q_0 & Q_1 & \cdots & Q_{n-1} \\ \vdots & \vdots &\ddots & \vdots\\Q_{-n+2} & Q_{-n+3} & \ldots & Q_1}, \nonumber\\ \nonumber \\
c^{xy}_{i,i+n} &= c^{yx}_{i,i+n} = c^{xz}_{i,i+n} = c^{zx}_{i,i+n} = c^{yz}_{i,i+n} = c^{zy}_{i,i+n} = 0,
\label{eqn: corrs}
\end{align}
as given in Ref.~\cite{mukherjee2018geometric}, and generalizing Refs.~\cite{pfeuty1970one, perk2009new} to finite temperature.

\subsection{Zero temperature numerical solution}\label{subsec: numerical}
We use MPS methods implemented in open source code openMPS~\footnote{Matrix product state open source code. http://sourceforge.net/projects/openmps/.} to find the ground state of the Ising model for finite systems with open boundary conditions~\cite{jaschke2017open}. MPS methods are highly versatile and applicable to a wide variety of interacting many-body systems, with the main constraint being the amount of entanglement allowed~\cite{schollwock2005density}. In essence, the algorithms perform data compression on the state of a many-body system by using a series of Schmidt decompositions on bipartitions of the lattice, followed by the truncation of a number of highly entangled states. OpenMPS utilizes a variational ground state search, a standard approach described in Refs.~\cite{schollwock2005density, jaschke2017open}. In our simulations, we set the variance tolerance $\expect{\hH_{\rm TIM}^2} - \expect{\hH_{\rm TIM}}^2 < 10^{-10}J^2L$, and study systems with size $L$ ranging from $10$ to $100$. The resulting ground state is well-converged.

\section{Network analysis for the TIM}\label{sec: network analysis for TIM}
In this section, we discuss the network structure of the TIM, with each subsection focusing on networks weighted by a different correlation measure.
\subsection{Von Neumann mutual information network}\label{subsec: TIM_MI}
\begin{figure}[t]
\makebox[-10cm]{\includegraphics[width=1.0\columnwidth]{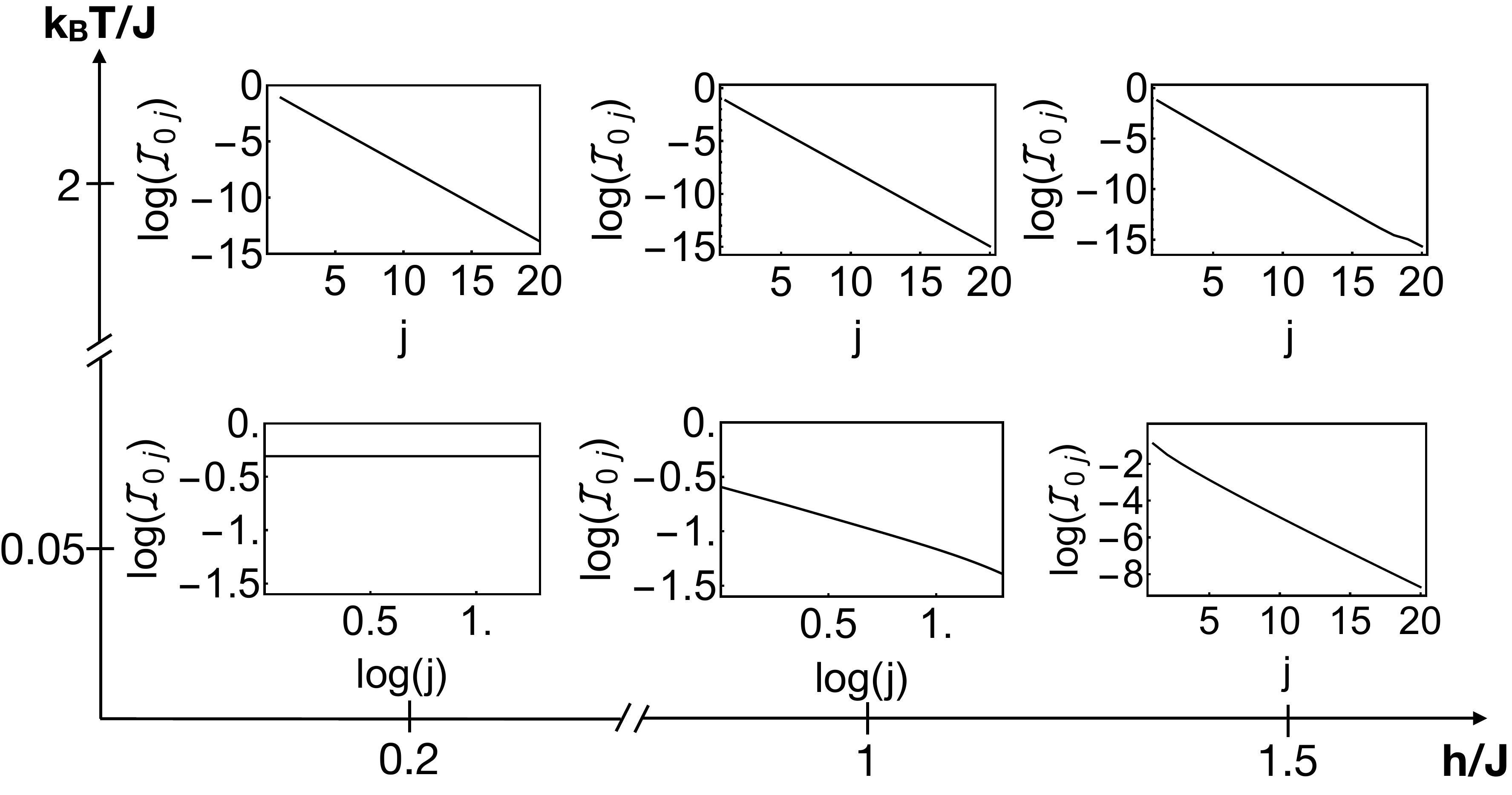}}
\caption{Von Neumann mutual information versus distance for the TIM, at different temperatures and transverse field couplings. The mutual information is nearly uniform when $\{h,k_BT\}\ll J$, as shown in the plot at the lower left corner. The mutual information algebraically decays near the quantum critical point, $k_BT\ll J\approx h$, as shown in the log-log plot in the lower middle. The mutual information exponentially decays in all other regimes, as shown in the remaining four log-linear plots.}
\label{fig: MIplots}
\end{figure}
\begin{figure}[t]
\centering
\includegraphics[width=1.0\columnwidth]{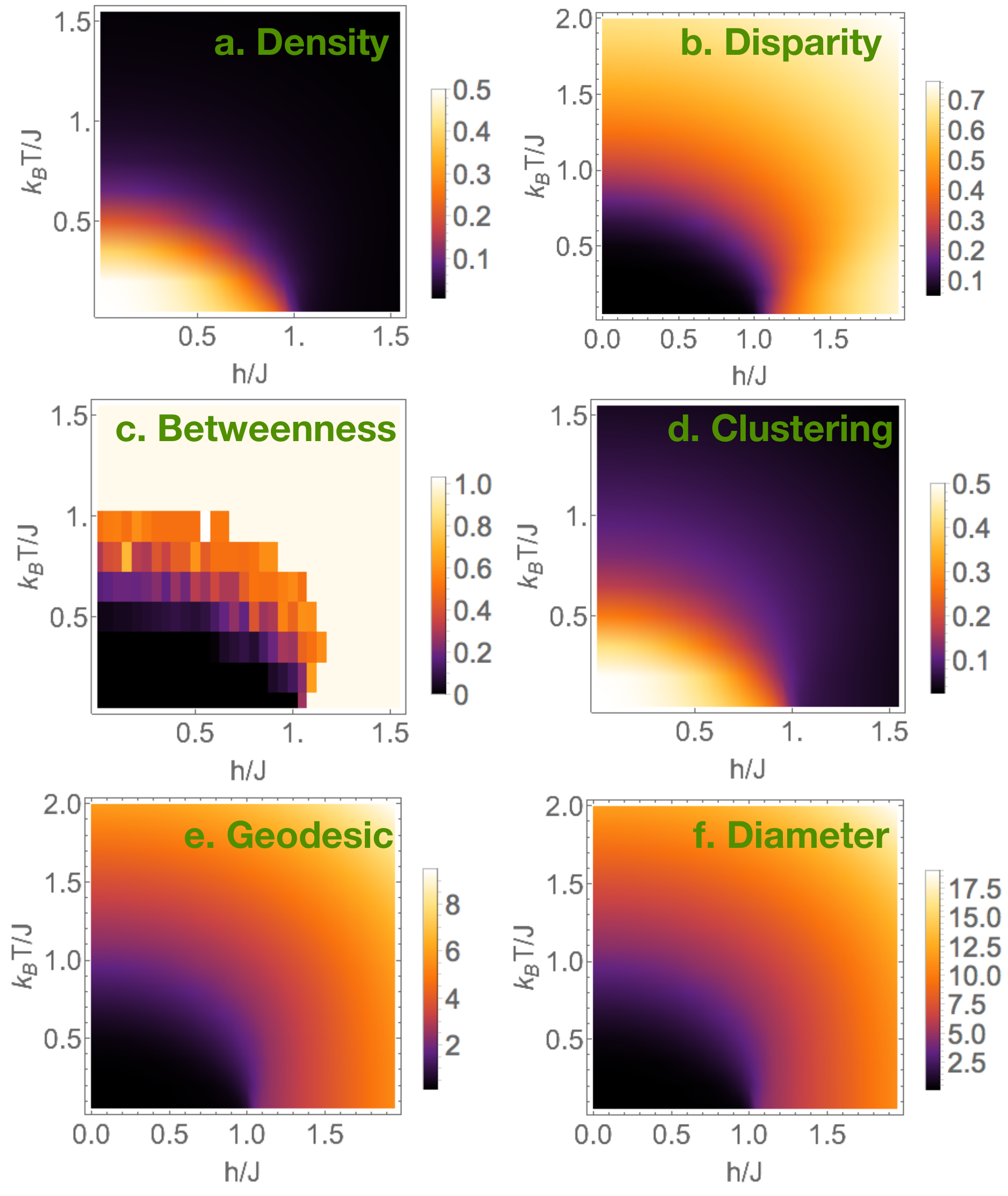}
\caption{(Color online) Network measures of the von Neumann mutual information network for the TIM in the thermodynamic limit, as a function of temperature and magnetic field. (a) Density of links connected to a node in the network. (b) Disparity of a node. (c) Normalized betweenness-centrality $\frac{B_i}{N^2}$ of a node $i$. (d) Clustering coefficient of the network. (e) Normalized average geodesic distance $\frac{\overline{D}}{N}$ between nodes. (f) Normalized diameter $\frac{D_{\rm max}}{N}$ of the network. The distance between two nodes across a link is defined as the inverse of the mutual information between them. The gradients of all network measures are observed to have an extremum at the quantum phase transition at $T=0,\ h=J$.}
\label{fig: TIM_MI}
\end{figure}
\begin{figure}[t]
\centering
\includegraphics[width=1.0\columnwidth]{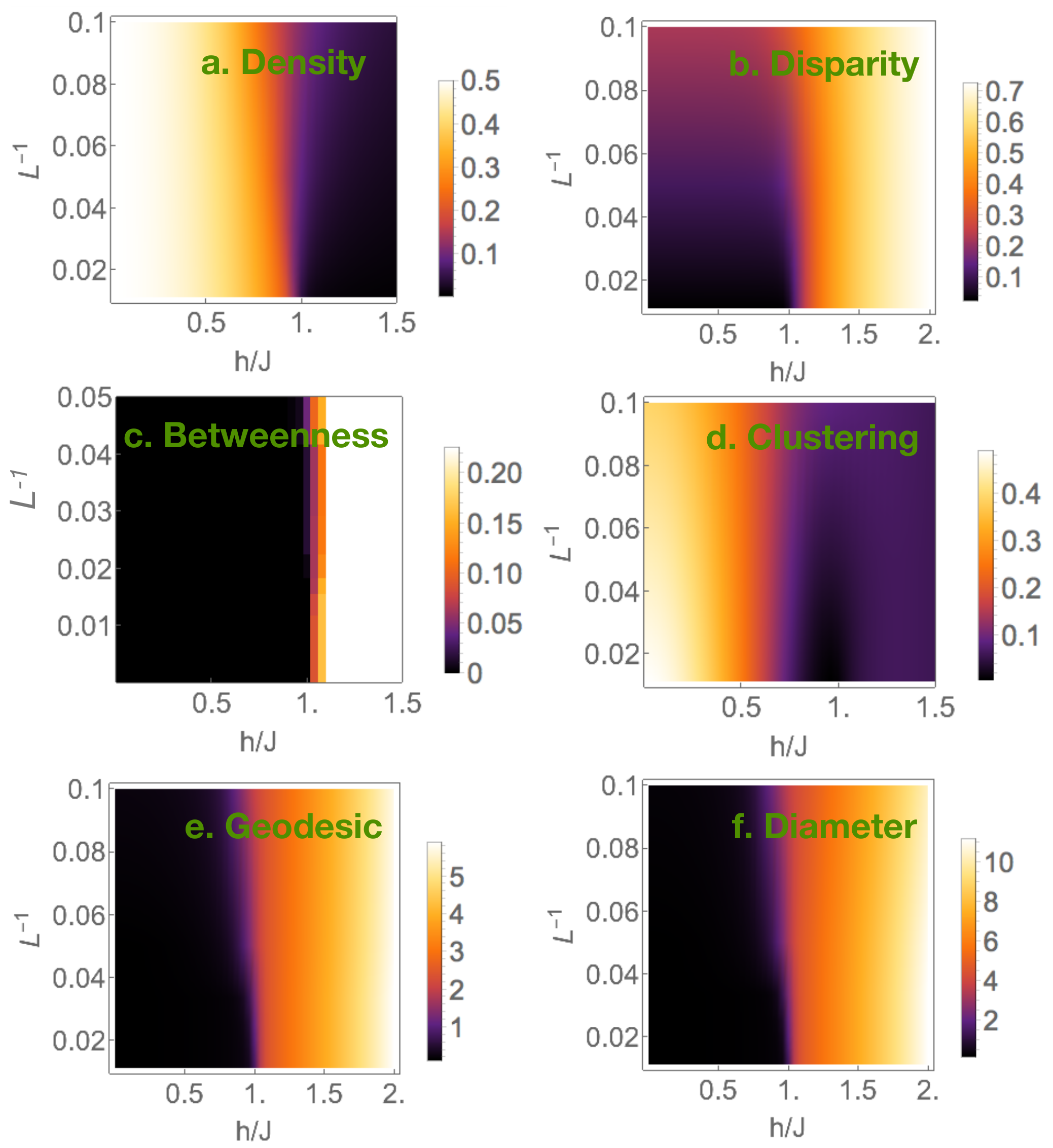}
\caption{(Color online) Network measures of the von Neumann mutual information network for the TIM at zero temperature, as a function of inverse system size and magnetic field. (a)-(f) plot the same quantities as in Fig.~\ref{fig: TIM_MI}, with inverse system size instead of temperature on the vertical axis.}
\label{fig: TIM_MI_openMPS}
\end{figure}

We calculate the von Neumann mutual information between two spins, as defined in Eq.~\eqref{eqn: MI}, and plot it as a function of their separation in Fig.~\ref{fig: MIplots}. We find that the mutual information is nearly uniform with distance when $\{h,k_BT\}\ll J$ (bottom left panel), indicating a long-range ordered phase. The mutual information does decay with distance, but very slowly, with a large correlation length $\xi\sim\mathcal{O}\left(\frac{-1}{\ln\tanh\beta J}\right) \sim \mathcal{O}\left(e^{\beta J}\right)$. The mutual information algebraically decays with distance when $k_BT\ll h\approx J$ (bottom middle panel), indicating the presence of a quantum critical point in the vicinity. It exponentially decays with distance in all other regimes, indicating a disordered phase of the spins.

The physics is further elucidated by the adjacency network built from the mutual information. Figure~\ref{fig: TIM_MI} shows all the network measures -- the density, disparity, betweenness-centrality, clustering coefficient, average geodesic distance, and diameter  -- of this weighted mutual information network, as a function of magnetic field and temperature.

In the ferromagnet (lower left panel in Fig.~\ref{fig: MIplots}), the nearly uniform spatial structure of the mutual information  yields a small disparity, and a large density and clustering coefficient. Nearly all geodesic paths between nodes are direct paths across one link. Hence, the betweenness-centrality, average geodesic distance, and diameter are all small. [Calculating the betweenness-centrality on a large network is computationally expensive. Therefore whenever $N>20$, we calculated the betweenness-centrality for a small region in the centre of the network.]

\begin{figure}[t]
\begin{tabular}{c}
\includegraphics[width=0.7\columnwidth]{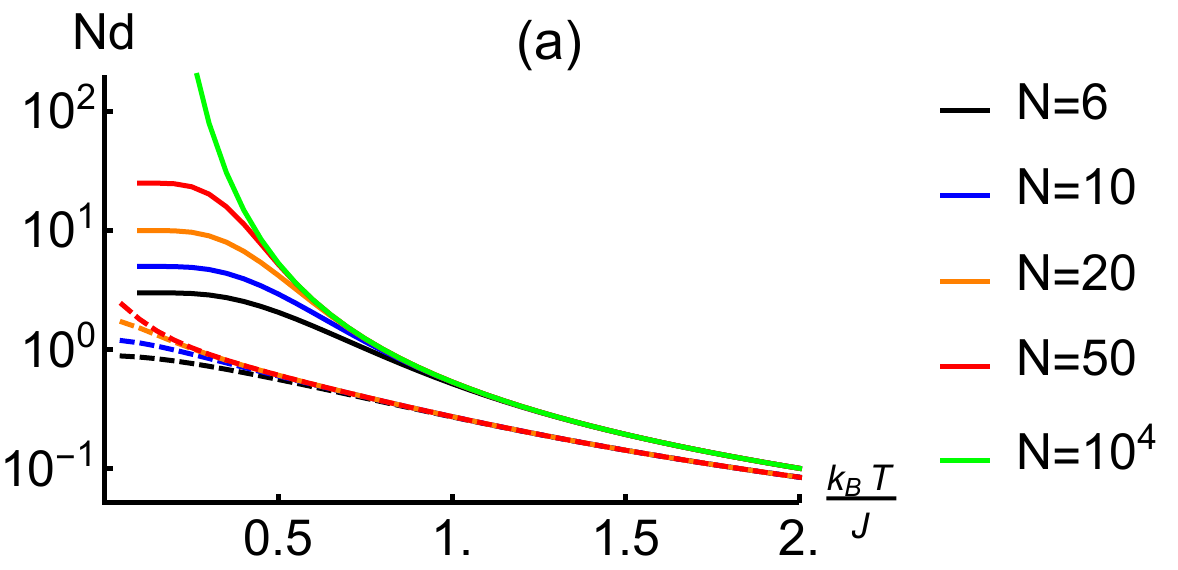} \\ \includegraphics[width=0.7\columnwidth]{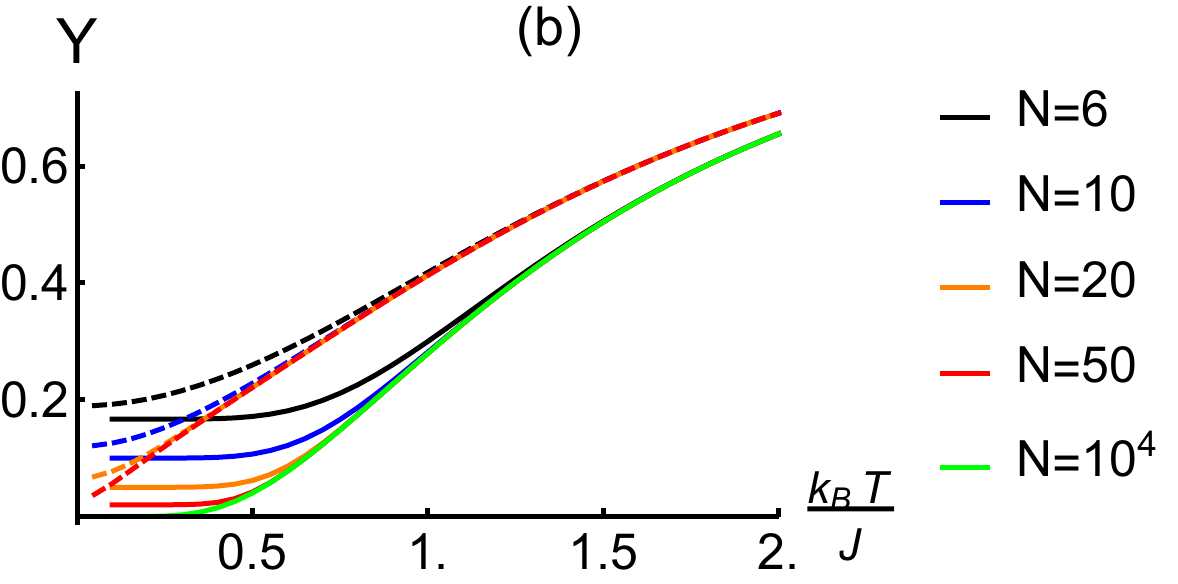}
\end{tabular}
\caption{(Color online) (a) The strength $\lambda=Nd$, and (b) disparity $Y$ of the von Neumann mutual information network, for different network sizes $N$. Solid lines: $h=0$, and dashed lines: $h=J$. At $T=0$, the strength diverges as $\lambda\sim\mathcal{O}(N)$, therefore the density is nonzero. For all $T>0$, the strength converges to a finite value as $N$ is increased, therefore the density for a network in the thermodynamic limit is $d=0$. Unlike the density, all other network measures converge to a finite function of temperature as the network grows in size.}
\label{fig: finite size scaling}
\end{figure}

In the paramagnet (bottom right and top panels in Fig.~\ref{fig: MIplots}), the mutual information decays exponentially with distance, resulting in a higher disparity, betweenness-centrality, average geodesic distance, and diameter, and a smaller clustering. As we discuss later in Fig.~\ref{fig: finite size scaling}, the density in the thermodynamic limit is zero everywhere in this phase. The nonzero density in Fig.~\ref{fig: MIplots}a for $T>0$ or $h>J$ is an artifact of working with a network of a finite size, $N\sim\mathcal{O}(100)$. The density converges to zero in this phase for network sizes $N>\xi$.

The most noticeable feature about the network measures is that they all change sharply across the phase transition from the ferromagnet to the paramagnet at $T=0, h=J$. All their gradients are observed to have extrema at the transition.

The network measures also provide information beyond the standard quantities -- correlation length and critical exponents -- that are used to characterize the phase transition. For example, in the quantum critical fan region near the critical point at $T=0, h=J$, the network structure differs from that in either the low-temperature disordered or ordered phase. In the critical fan, the density and clustering coefficient appear closer to those of the paramagnetic phase, while the other network measures resemble the ferromagnetic phase. The width of the fan also appears to be different for the different network measures.

We address finite size effects on two fronts. First, in Fig.~\ref{fig: TIM_MI_openMPS}, we show finite-size effects at zero temperature using MPS for all complex network measures. The critical point is still clearly evident even for tens of sites, and moves toward $h/J=1$ from below, becoming sharper as the system size increases. For a more detailed study of finite size scaling effects, see Ref.~\cite{valdez2017quantifying}.

Second, it should be noted that in Fig.~\ref{fig: TIM_MI}, we use analytic expressions for the reduced density matrices, that are valid in the thermodynamic limit $L=\infty$. However, we calculate network measures for adjacency networks truncated to $N\sim\mathcal{O}(100)$ nodes, assuming that the correlations have sufficiently decayed when the separation between spins is $\mathcal{O}(100)$. To analyze the convergence of our network measures as $N$ increases, we plot the strength $\lambda=Nd$ and disparity $Y$ for different network sizes in Fig.~\ref{fig: finite size scaling}. We find that the disparity converges to a finite value for all $T, h$ and $N$. The strength converges to a finite value for $T>0$ or $h\leq J$ and large enough $N\gtrsim50$, implying that the density is $d=0$ in the thermodynamic limit of the network, $N\rightarrow\infty$. However, when $T=0,\ h<J$ and for large $N$, the strength diverges as $Nd\sim\mathcal{O}(N)$, yielding a nonzero density. As a result, the density undergoes a discontinuous jump from $0$ to a finite value as the system is tuned from $T\rightarrow0^+$ to $T=0$ (which corresponds to tuning from the paramagnetic to the ferromagnetic phase). Therefore, the density of the weighted mutual information network is a good order parameter for the ferromagnetic phase.  All the other network measures converge to finite functions of temperature at a large enough $N\sim50$.

\subsection{R\'enyi mutual information networks}\label{subsec: TIM_MIRenyi}
\begin{figure}[t]
\centering
\includegraphics[width=1.0\columnwidth]{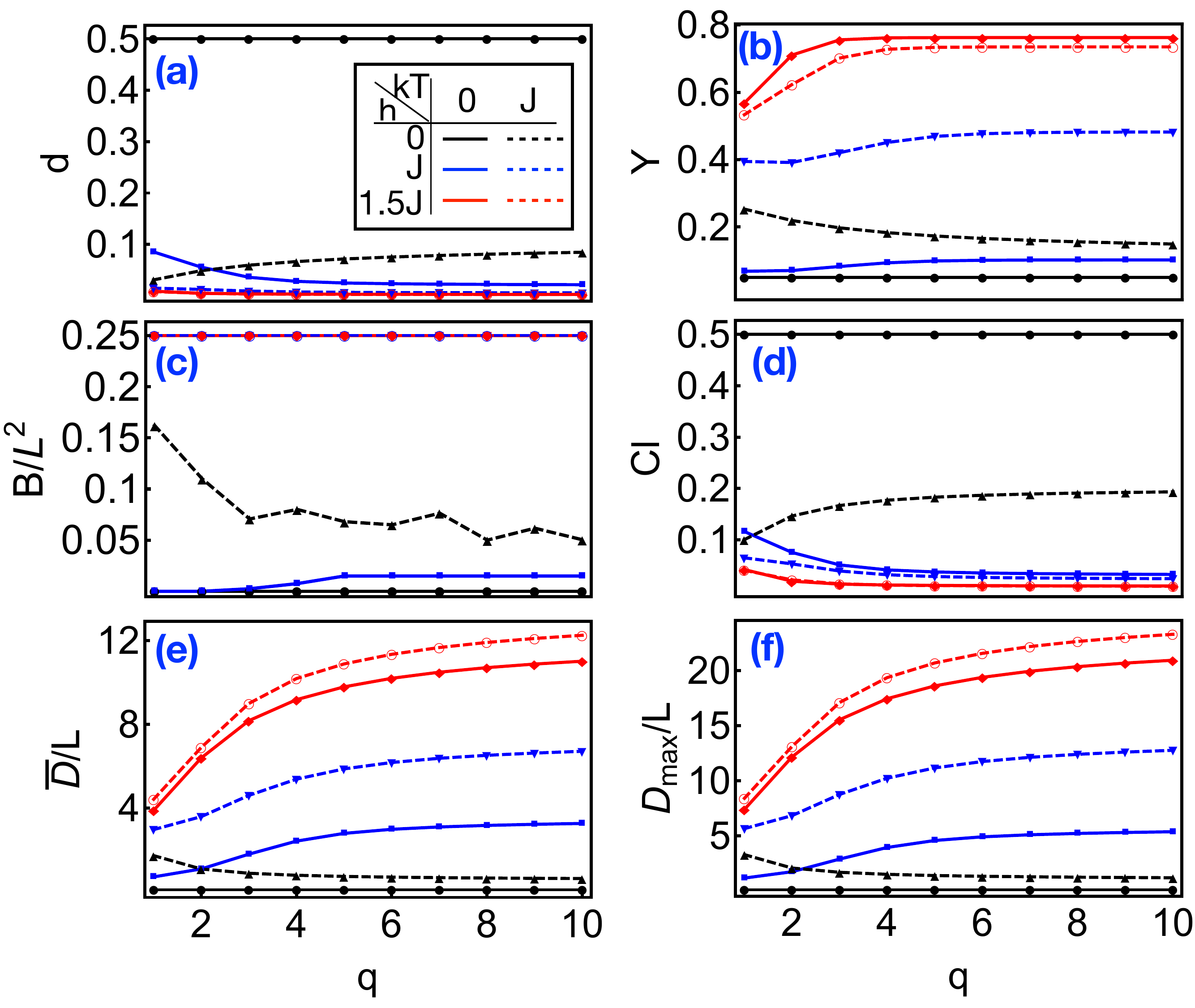}
\caption{(Color online) Network measures of the R\'enyi mutual information network as a function of R\'enyi order $q$, for the TIM in the thermodynamic limit, at different magnetic fields and temperatures. (a)-(f) plot the same quantities as Fig.~\ref{fig: TIM_MI} versus R\'enyi order $q$, at specific magnetic fields and temperatures specified in the inset in (a).
}
\label{fig: TIM_MIRenyi}
\end{figure}
\begin{figure}[t]
\centering
\includegraphics[width=1.0\columnwidth]{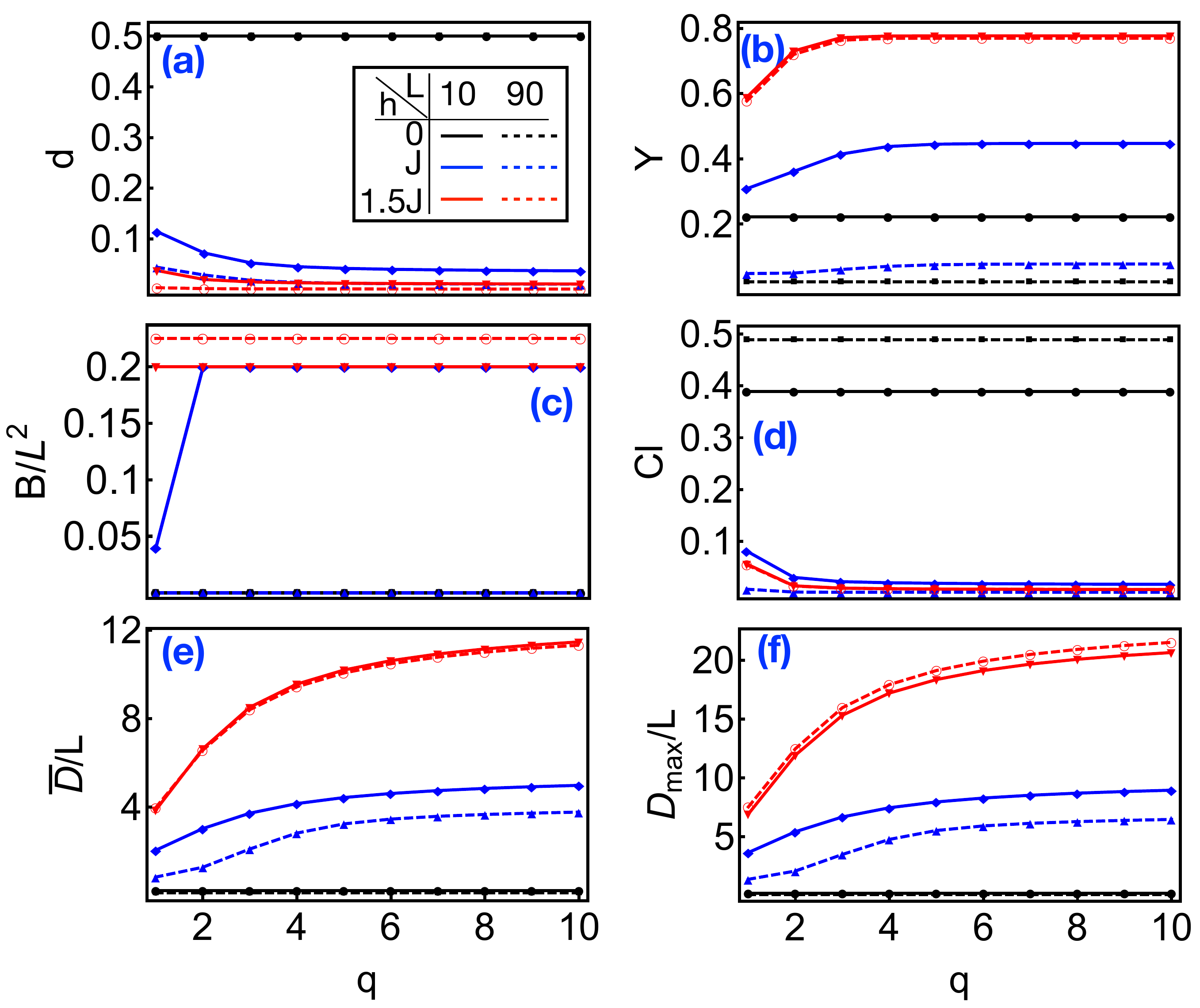}
\caption{(Color online) Network measures of the R\'enyi mutual information network as a function of R\'enyi order $q$, for the TIM at zero temperature, at different magnetic fields and system sizes. (a)-(f) plot the same quantities as Fig.~\ref{fig: TIM_MI} versus R\'enyi order $q$, at specific magnetic fields and system sizes specified in the inset in (a).
}
\label{fig: TIM_MIRenyi_openMPS}
\end{figure}
R\'enyi mutual informations also provide useful information about a system. The von Neumann information is a special case of the R\'enyi information: $\mathcal{I}_{ij} = \lim_{q\rightarrow1}\mathcal{I}_{ij}^q$. While the von Neumann mutual information measures the sum of the log of the eigenvalues of the reduced density matrix weighted equally, the R\'enyi generalization skews the weights towards the largest eigenvalues. Like the von Neumann information, the R\'enyi mutual information can be readily extracted from tomography measurements~\cite{roos2004bell, haffner2005scalable, cramer2010efficient, lanyon2017efficient} of reduced density matrices, using Eqs.~\eqref{eqn: rho1},~\eqref{eqn: rho2}, and~\eqref{eqn: Renyi}. Here, we calculate network properties of adjacency networks for the R\'enyi mutual information between spins.

Figure~\ref{fig: TIM_MIRenyi} shows all the network measures for the R\'enyi mutual information network at different temperatures and magnetic fields, as a function of R\'enyi order $q$. Figure~\ref{fig: TIM_MIRenyi_openMPS} shows all the network measures for the R\'enyi mutual information network at $T=0$ and different magnetic fields for systems with two different sizes, as a function of R\'enyi order $q$. Like the von Neumann mutual information, the R\'enyi mutual information network at any fixed order $q$ has a different structure in the ferromagnetic and paramagnetic phases and the critical fan region.

Deep in the ferromagnetic phase at $h=0,T=0$, the $q^{\rm th}$ order R\'enyi information between spins $i$ and $j$ is $\mathcal{I}^q_{ij}=0.5$. As a result, we observe in Fig.~\ref{fig: TIM_MIRenyi} that the density and clustering coefficient are $0.5$, and all other network measures are zero deep in this phase.

The R\'enyi information networks are observed to undergo a sharp change at the quantum phase transition from the ferromagnetic to the paramagnetic phase at $T=0,\ h=J$. In the paramagnet, the R\'enyi information decays rapidly with separation. We observe in Fig.~\ref{fig: TIM_MIRenyi} that the density and clustering coefficient are $0$, and all other network measures are nonzero in this phase. In Fig.~\ref{fig: TIM_MIRenyi_openMPS}, we observe that all network measures approach their thermodynamic limits as $L$ increases.

We explore the $L$-dependence of the network measures in more detail in Fig.~\ref{fig: TIM_MIRenyi vs L} in the Appendix.

\subsection{Spin-spin correlation networks}\label{subsec: TIM_corrs}
\begin{figure}[t]
\centering
\includegraphics[width=1.0\columnwidth]{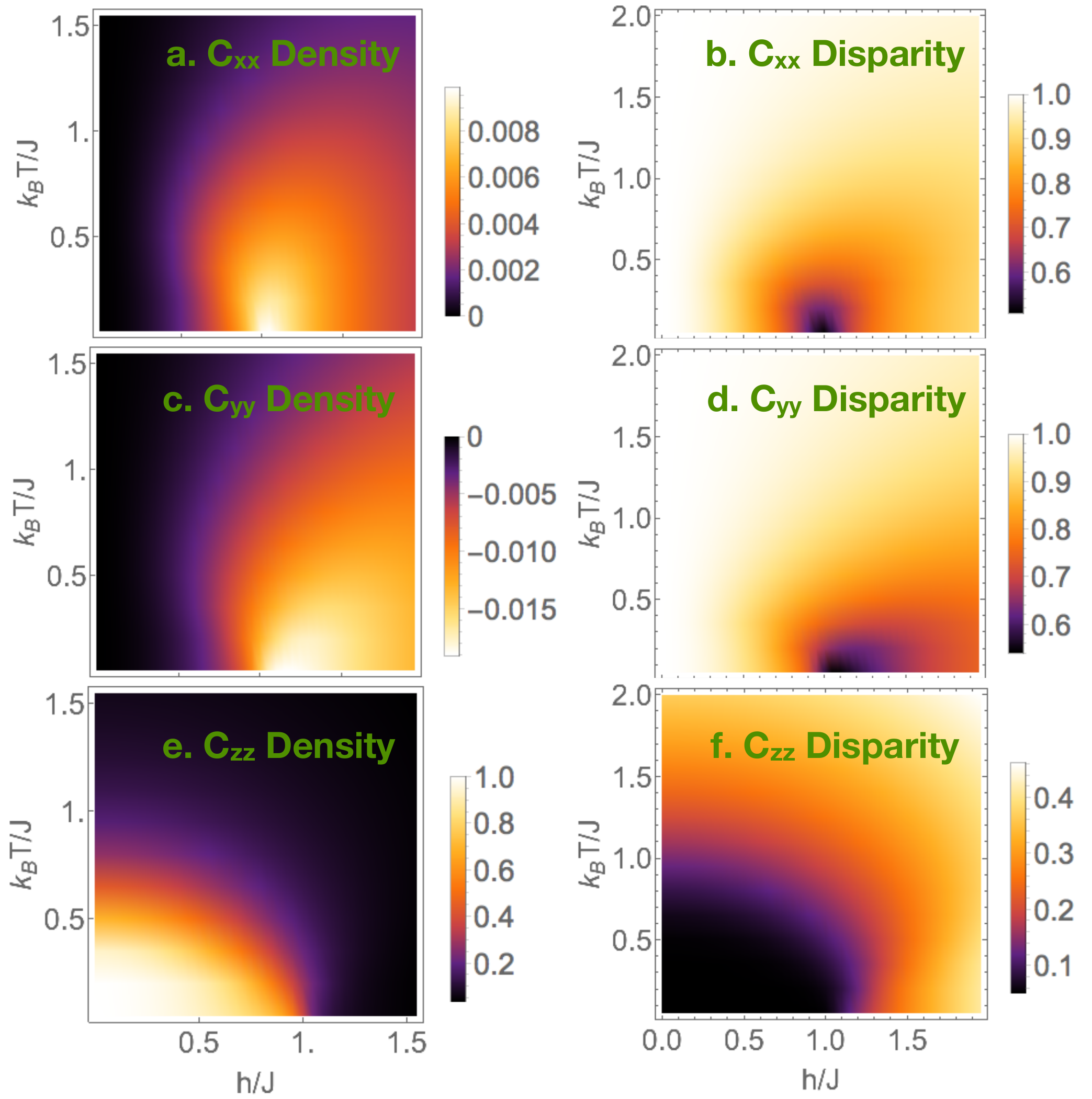}
\caption{(Color online) Network measures of the spin-spin correlation networks for the TIM. Left panels: Density of links connected to a node in the adjacency network for different spin correlations. Right panels: Disparity of a node in these adjacency networks.}
\label{fig: TIM_corrs}
\end{figure}
In this section, we consider networks weighted by the connected correlations $C^{\mu\nu}_{ij}=\expect{\hsigma^\mu_i\hsigma^\nu_j}-\expect{\hsigma^\mu_i}\expect{\hsigma^\nu_j}$. From Eq.~\eqref{eqn: corrs}, only the diagonal components $C_{ij}^{\mu\mu}$ are nonzero. Figure~\ref{fig: TIM_corrs} shows the density and disparity of these networks, as a function of magnetic field and temperature.

We find that the $C_{zz}$ network shares features similar to the mutual information network. This is expected, because $C_{zz}$ is the dominant correlation. Again, the quantum phase transition at $T=0,\ h=J$ distinctly stands out: the gradients of all the network measures are observed to have an extremum at this phase transition.

The networks built from $C_{xx}$ and $C_{yy}$ also have intriguing characteristics. Unlike the mutual information and $C_{zz}$, which are uniform in the ferromagnet and exponentially decaying with distance in the paramagnet, $C_{xx}$ and $C_{yy}$ exponentially decay in both the paramagnet and ferromagnet. They algebraically decay with distance near the phase transition. The density of the $C_{xx}$ and $C_{yy}$ networks is observed to have a maximum, and the disparity of both networks observed to have a minimum at the phase transition.

\subsection{Concurrence and negativity networks}\label{subsec: TIM_conneg}
Concurrence [Eq.~\eqref{eqn: con}] is a non-negative entanglement monotone that indicates if two spins are entangled. Negativity [Eq.~\eqref{eqn: neg}] is a complete entanglement witness that also indicates entanglement between spins. For entangled spins, their concurrence is positive and their negativity negative. For unentangled spins, both concurrence and negativity are zero. For the TIM, the concurrence and negativity between two spins $i$ and $j$ are (see Appendices)
\begin{equation}
\mathcal{C}_{ij} = {\rm max}\left(0,-\frac{1}{2}\left(1-c^{xx}_{ij}+c^{yy}_{ij}-c^{zz}_{ij}\right)\right),
\end{equation}
\begin{equation}
\mathcal{N}_{ij} = {\rm min}\left(0,\frac{1}{4}\left(1-c^{xx}_{ij}+c^{yy}_{ij}-c^{zz}_{ij}\right)\right) = -\frac{1}{2}\mathcal{C}_{ij},
\end{equation}
with $c_{ij}^{\mu\nu}$ given by Eq.~\eqref{eqn: corrs}. For this model, the concurrence and negativity networks predominantly have only nearest-neighbor connections. Therefore, the densities of the networks in the thermodynamic limit are zero everywhere in the phase diagram.

\begin{figure}[t]
\centering
\includegraphics[width=1.0\columnwidth]{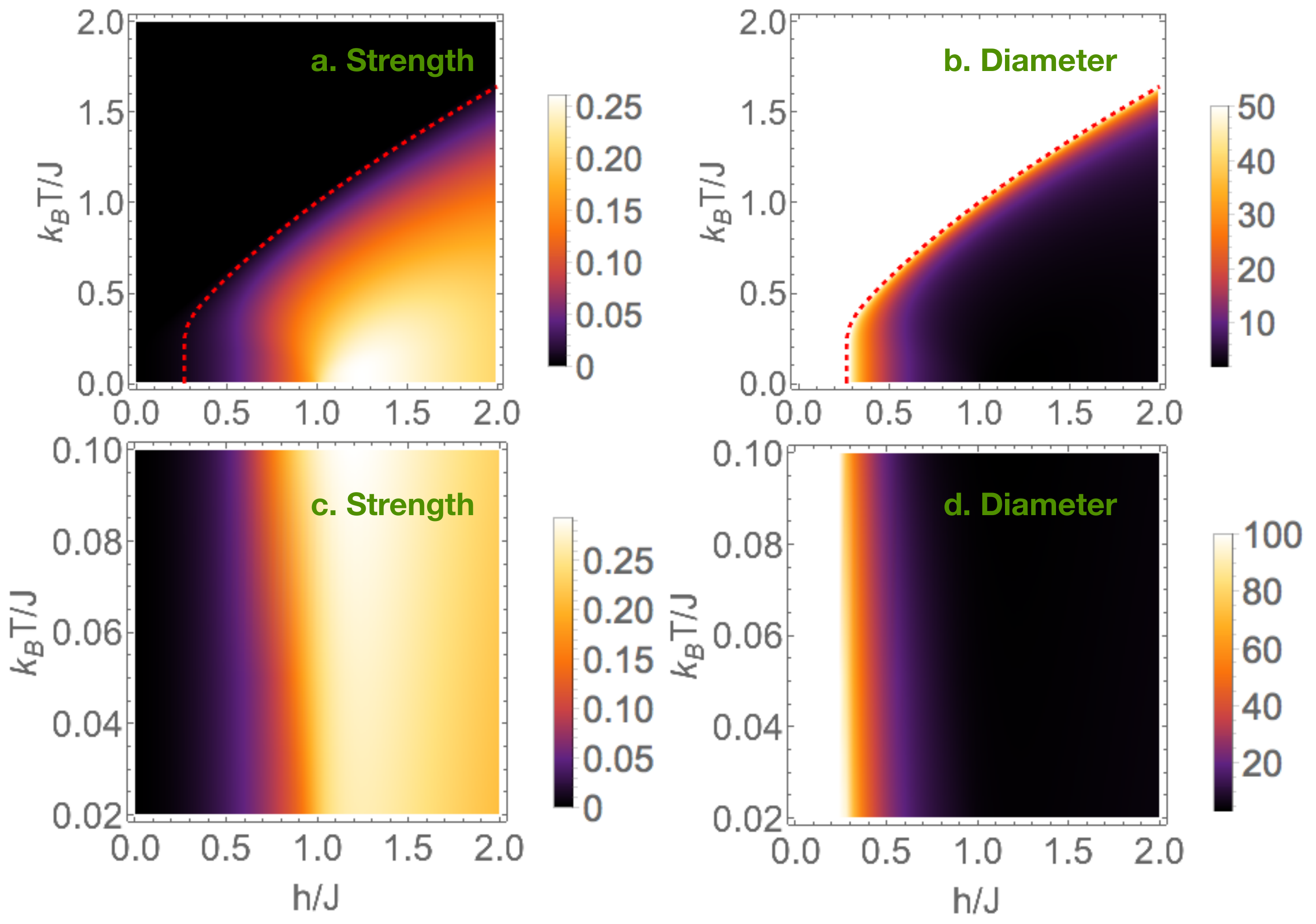}
\caption{Network measures of the concurrence network for the TIM. The top panels show (a) the strength of a node ($=N\times$ density) in the network, and the normalized diameter $\frac{D_{\rm max}}{N}$, at a finite temperature (vertical axis) for a system in the thermodynamic limit. The bottom panels show the same measures at zero temperature for systems of lengths ranging from $10$ to $50$ sites (vertical axis). Above and to the left of the dotted line in all four panels, the concurrence is zero for all the links in the network, the strength is zero and the diameter of the network is infinite. The gradient of the density is observed to have a maximum at the quantum phase transition at $T=0,\ h=J$ in the thermodynamic limit.}
\label{fig: TIM_con}
\end{figure}

We plot the strength of a node ($\lambda=Ld$), and the diameter of the concurrence network in Fig.~\ref{fig: TIM_con}. We find that the strength is nonzero in a region around $T=0,\ h=J$. The gradient of the strength is observed to have an extremum at $T=0,\ h=J$. The concurrence between all pairs of spins is zero above a critical temperature, indicated by the dotted line in Fig.~\ref{fig: TIM_con}. This critical temperature is not associated with any phase transition, but with the sudden death of entanglement between spins (for other examples of entanglement sudden death, see e.g Ref.~\cite{yu2009sudden}). Above this critical temperature, the concurrence network is trivial and has all link weights as zero. The disparity is nearly $1$ everywhere below the dotted line. Similarly, the normalized betweenness-centrality is always nearly $\frac{B}{L^2}=\frac{1}{4}$, and the clustering coefficient is nearly $0$. The diameter and average geodesic distance are finite and related as $\overline{D}\approx D_{\rm max}/2$ below the dotted line. The diameter and average geodesic distance are infinite above this line. The structure of the negativity network is identical to that of the concurrence network.

\section{Conclusions}\label{sec: conclusions}
The correlations between particles in an interacting quantum system naturally form a weighted network. Characterizing quantum systems via network properties of their correlation networks is a new paradigm for exploring and visualizing quantum systems. Since correlations between particles are measurable in experiments~\cite{roos2004bell, haffner2005scalable, cramer2010efficient, lanyon2017efficient, parsons2016site, cheuk2015quantum, edge2015imaging, omran2015microscopic, haller2015single, yamamoto2016ytterbium, richerme2014non, zhang2017observation, zeiher2016many, greif2015formation, bernien2017probing}, network analysis of the correlations will be a useful tool to understand the underlying physics of the system. We have shown that networks for various correlation measures, such as spin-spin correlations, von Neumann and R\'enyi mutual information, concurrence, and negativity exhibit emergent complexity even for simple Hamiltonians such as the one-dimensional transverse field Ising model. We used network analysis tools to characterize the complexity of these networks, and showed that the network measures provide a wealth of information about the system throughout the entire phase diagram, above and beyond the usual quantities such as the correlation length and critical exponents. For example, all network measures for most correlation networks had entirely different signatures in the different phases of the Ising system, and exhibited distinct sharp features at the quantum phase transition from the ferromagnetic to the paramagnetic phase. The network measures also show intriguing features in the critical fan region near the phase transition, where the network structure is different from both the ordered and disordered phases. 

We predict that this new paradigm  of visualizing a quantum system as a network will have important implications for future experimental as well as theoretical work. For example, we have already argued and demonstrated that some network measures effectively play the role of an order parameter, and all network measures are effective at identifying equilibrium phases and phase transitions. We also expect the correlation networks to exhibit particularly interesting behavior with time in quench or ramp experiments commonly performed with cold atoms or trapped ions, since the propagation of correlations after a quench or ramp may be efficiently visualized using changes in the correlation network's structure.

\section*{acknowledgment}
This material is based upon work supported with funds from the Welch Foundation, grant no. C-1872. KRAH and LDC thank the Aspen Center for Physics, which is supported by the National Science Foundation grant PHY-1066293, for its hospitality while part of this work was performed. LDC and MAV acknowledge support by the National Science Foundation under the grants PHY1520915, and OAC-1740130, the Air Force Office of Scientific Research under grant FA9550-14-1-0287, and the U.K. Engineering and Physical Sciences Research Council through the ``Quantum Science with Ultracold Molecules'' Programme (Grant No. EP/P01058X/1).  The calculations were carried out in part using the high performance computing resources provided by the Golden Energy Computing Organization at the Colorado School of Mines.

\appendix

\begin{figure}[t]
\centering
\includegraphics[width=1.0\columnwidth]{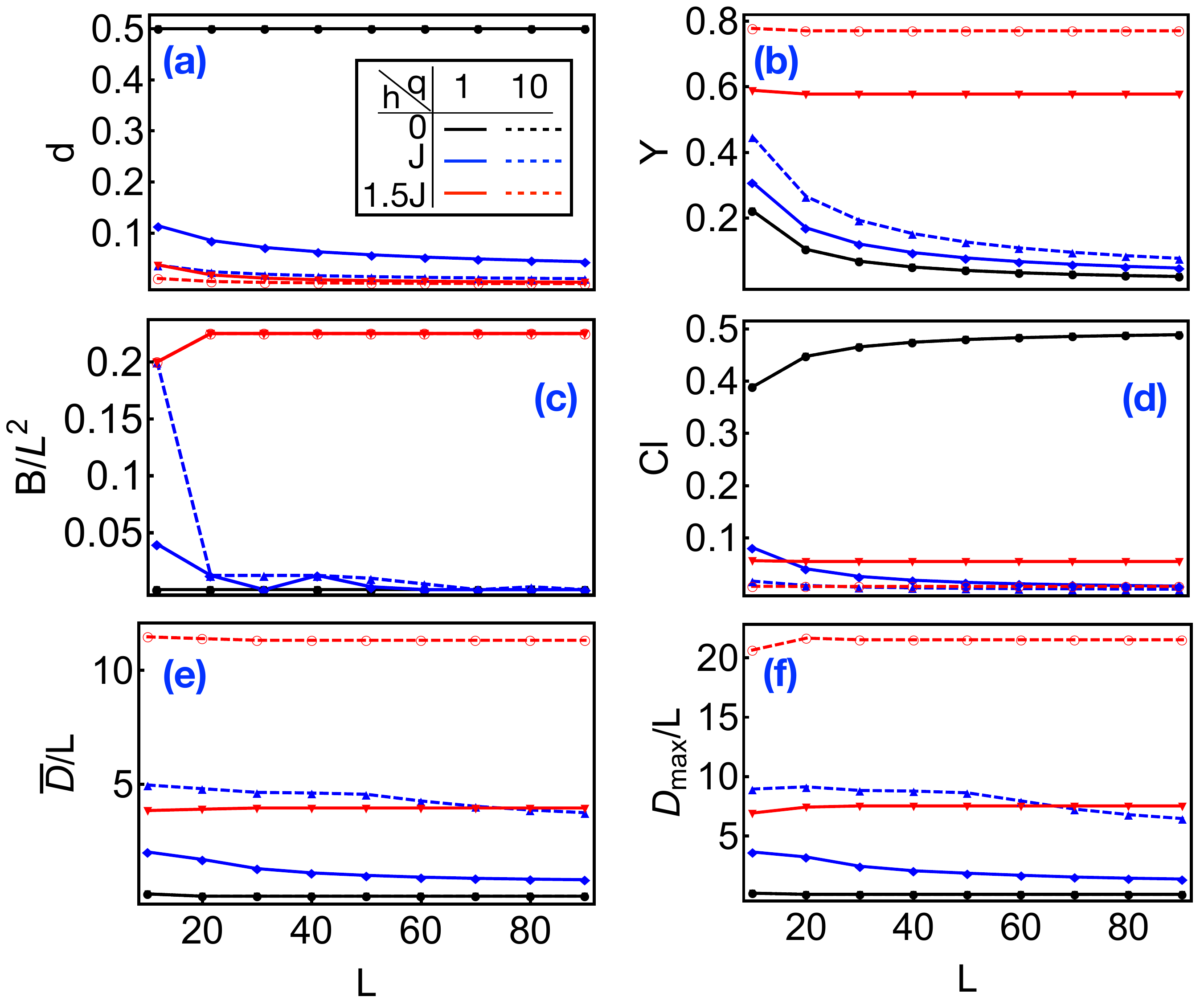}
\caption{Network measures of the R\'enyi mutual information network as a function of system size $L$, for the TIM at $T=0$, at different magnetic fields and R\'enyi order $q$. (a)-(f) plot the same quantities as Fig.~\ref{fig: TIM_MI} versus system size $L$, at magnetic fields and R\'enyi order specified in the inset in (a).}
\label{fig: TIM_MIRenyi vs L}
\end{figure}

\section{Concurrence between spins}
The two-spin reduced density matrix in the TIM is
\begin{equation}\label{eqn: explicit_rho2}
\rho_{ij}^{(2)} = \frac{1}{4}\left(\begin{array}{cccc}
1+c^{zz}_{ij} & m^x_j & m^x_i & c^{xx}_{ij}-c^{yy}_{ij} \\ m^x_j & 1-c^{zz}_{ij} & c^{xx}_{ij}+c^{yy}_{ij} & m^x_i \\ m^x_i & c^{xx}_{ij}+c^{yy}_{ij} & 1-c^{zz}_{ij} & m^x_j \\ c^{xx}_{ij}-c^{yy}_{ij} & m^x_i & m^x_j & 1+c^{zz}_{ij}
\end{array}\right).
\end{equation}
Due to translational invariance, $m^x_i = m^x_j = m^x$. Let $R(\rho) = \sqrt{\sqrt{\rho}\tilde{\rho}\sqrt{\rho}}$, with $\tilde{\rho} = \rho(m^x\rightarrow-m^x)$ the spin-flipped density matrix. The eigenvalues $\lambda$ of $R(\rho)$ satisfy
\begin{equation}
\det(R^2(\rho)-\lambda^2) = 0.
\end{equation}
Multiplying inside the determinant by $\sqrt{\rho^{-1}}$ on the left and $\sqrt{\rho}$ on the right, we find that $\lambda^2$ is also an eigenvalue of $\tilde{\rho}\rho$. Therefore, all eigenvalues $\lambda$ of $R(\rho)$ are eigenvalues of $R'(\rho) = \sqrt{\tilde{\rho}\rho}$ as well. The eigenvalues of $R'(\rho_{ij}^{(2)})$ are (in decreasing order),
\begin{align}
&\lambda_1 = \frac{1}{4}\left(\sqrt{(1+c^{xx}_{ij})^2-4(m^x)^2}+c^{zz}_{ij}-c^{yy}_{ij}\right),\nonumber\\
&\lambda_2 = \frac{1}{4}\left(1-c^{xx}_{ij}+c^{yy}_{ij}+c^{zz}_{ij}\right),\nonumber\\
&\lambda_3 = \frac{1}{4}\left(1-c^{xx}_{ij}-c^{yy}_{ij}-c^{zz}_{ij}\right),\nonumber\\
&\lambda_4 = \frac{1}{4}\left(\sqrt{(1+c^{xx}_{ij})^2-4(m^x)^2}-c^{zz}_{ij}+c^{yy}_{ij}\right).
\end{align}
Therefore, the concurrence between two spins is
\begin{align}
\mathcal{C} &= {\rm max}\left(0,\lambda_1-\lambda_2-\lambda_3-\lambda_4\right)\nonumber\\
 &= {\rm max}\left(0,-\frac{1}{2}\left(1-c^{xx}_{ij}+c^{yy}_{ij}-c^{zz}_{ij}\right)\right).
\end{align}
\\ \\
\section{Negativity between spins}
Let $\tilde{\tilde{\rho}}^{(2)}_{ij} = \left(\mathbf{1}\otimes {\rm T}\right)\rho_{ij}^{(2)}$. $\tilde{\tilde{\rho}}^{(2)}_{ij}$ has a form identical to Eq.~\eqref{eqn: explicit_rho2}, with $c^{yy}_{ij}\rightarrow-c^{yy}_{ij}$. The eigenvalues of $\tilde{\tilde{\rho}}^{(2)}_{ij}$ are
\begin{align}
&\lambda'_1 = \frac{1}{4}\left(1-c^{xx}_{ij}+c^{yy}_{ij}-c^{zz}_{ij}\right),\nonumber\\
&\lambda'_2 = \frac{1}{4}\left(1-c^{xx}_{ij}-c^{yy}_{ij}+c^{zz}_{ij}\right),\nonumber\\
&\lambda'_3 = \frac{1}{4}\left(1+c^{xx}_{ij}-\sqrt{(c^{yy}_{ij}+c^{zz}_{ij})^2+4(m^x)^2}\right),\nonumber\\
&\lambda'_4 = \frac{1}{4}\left(1+c^{xx}_{ij}+\sqrt{(c^{yy}_{ij}+c^{zz}_{ij})^2+4(m^x)^2}\right).
\end{align}
Of these eigenvalues, $\lambda'_{2,3,4}$ are always positive. Since $\lambda'_1+\lambda'_2+\lambda'_3+\lambda'_4=1$, we have
\begin{align}
\mathcal{N}_{ij} = \frac{\operatorname{Tr}\left|\tilde{\tilde{\rho}}_{ij}^{(2)}\right|-1}{2} = \frac{|\lambda'_1|-\lambda'_1}{2} = {\rm min}(0,\lambda'_1).
\end{align}

\section{Network measures for R\'enyi mutual information at different system sizes}
Figure~\ref{fig: TIM_MIRenyi vs L} plots the network measures for the R\'enyi information versus system size. All network measures converge to their thermodynamic values as $L$ increases.

In the thermodynamic limit in the ferromagnetic phase, the Renyi information $I_{ij}^q=0.5$ is uniform with distance. It follows that $d$ and Cl are $0.5$, and other network measures converge to zero.

In the thermodynamic limit in the paramagnetic phase, $I_{ij}^q$ decays rapidly with separation. Therefore $d$ and Cl converge to $0$. The geodesic path from an arbitrary spin $i$ to $j$ in the Renyi information network travels via all intervening spins $i+1, i+2, .. j-1$. Assuming translational invariance, the length of this geodesic path is $\frac{|i-j|}{I_{i,i+1}^q}$. It follows that $\frac{D_{\rm max}}{L} = \frac{1}{I^q_{i,i+1}}$, $\frac{\overline{D}}{L} = \frac{2}{I^q_{i,i+1}}$, and $\frac{B}{L^2} = \frac{1}{4}$, where $I^q_{i,i+1}$ is independent of $i$. In the critical region, $d$, $Y$, Cl, and $\frac{B}{L^2}$ converge to $0$ at large $L$, while $\frac{D_{\rm max}}{L}$ and $\frac{\overline{D}}{L}$ saturate to a nonzero value.

\bibliography{Bibfile}
\end{document}